\documentclass[conference]{IEEEtran}
\IEEEoverridecommandlockouts
\usepackage{cite}
\usepackage{amsmath,amssymb,amsfonts}
\usepackage{graphicx}
\usepackage{textcomp}
\usepackage{xcolor}
\usepackage{verbatim}
\usepackage{algorithm}
\usepackage{wasysym}
\usepackage[noend]{algpseudocode}
\def\BibTeX{{\rm B\kern-.05em{\sc i\kern-.025em b}\kern-.08em
    T\kern-.1667em\lower.7ex\hbox{E}\kern-.125emX}}
\begin{document}

\title{A near-autonomous and incremental intrusion detection system through active learning of known and unknown attacks
}

\makeatletter
\newcommand{\linebreakand}{%
  \end{@IEEEauthorhalign}
  \hfill\mbox{}\par
  \mbox{}\hfill\begin{@IEEEauthorhalign}
}
\makeatother

\author{\IEEEauthorblockN{1\textsuperscript{st} Lynda Boukela}
\IEEEauthorblockA{\textit{School of Computer Science and Engineering} \\
\textit{Nanjing University of Science and Technology}\\
Nanjing, China\\
lyndaboukela@njust.edu.cn}\\

\and

\IEEEauthorblockN{2\textsuperscript{nd} Gongxuan Zhang}
\IEEEauthorblockA{\textit{School of Computer Science and Engineering} \\
\textit{Nanjing University of Science and Technology}\\
Nanjing, China \\
gongxuan@njust.edu.cn}

\linebreakand
\IEEEauthorblockN{3\textsuperscript{rd} Meziane Yacoub}
\IEEEauthorblockA{\textit{CEDRIC lab} \\
\textit{Conservatoire National des Arts et M\'etiers}\\
Paris, France\\
meziane.yacoub@cnam.fr}

\and
\IEEEauthorblockN{4\textsuperscript{th} Samia Bouzefrane}
\IEEEauthorblockA{\textit{CEDRIC lab} \\
\textit{Conservatoire National des Arts et M\'etiers}\\
Paris, France\\
samia.bouzefrane@cnam.fr}

}

\maketitle

\begin{abstract}
Intrusion detection is a traditional practice of security experts, however, there are several issues which still need to be tackled. Therefore, in this paper, after highlighting these issues, we present an architecture for a hybrid Intrusion Detection System (IDS) for an adaptive and incremental detection of both known and unknown attacks. The IDS is composed of supervised and unsupervised modules, namely, a Deep Neural Network (DNN) and the K-Nearest Neighbors (KNN) algorithm, respectively. The proposed system is near-autonomous since the intervention of the expert is minimized through the active learning (AL) approach. A query strategy for the labeling process is presented, it aims at teaching the supervised module to detect unknown attacks and improve the detection of the already-known attacks. This teaching is achieved through sliding windows (SW) in an incremental fashion where the DNN is retrained when the data is available over time, thus rendering the IDS adaptive to cope with the evolutionary aspect of the network traffic. A set of experiments was conducted on the CICIDS2017 dataset in order to evaluate the performance of the IDS, promising results were obtained.
\end{abstract}


\section{Introduction}
In the recent years, an explosive growth of networks of all types has been noticed and the number of connected objects has been increasing rapidly. This results in users facing permanently a plethora of security threats. As an example, in the McAfee Labs Threats Report \cite{R1}, one can see that, in the first quarter of 2019, ransomware attacks grew by 118\%, new ransomware families were detected, and threat actors used innovative techniques.
 Intrusion Detection Systems (IDSes) are widely used as a reactive defense strategy. Researchers have explored widely machine learning (ML) and data mining (DM) techniques and they have adopted different detection strategies. Indeed, IDSes can be (i) signature-based, (ii) model-based, (iii) unsupervised, or (iv) hybrid \cite{R2.72}. Each of these detection strategies present some challenges, such as, the non ability to cope with unseen attacks, the important human intervention for data labeling and the need for a regular update due to the ever-evolution of the network traffic. Therefore, the intrusion detection methodology involves a constant trade-off between the ability of the system to cope with unknown attacks, its detection performance and the autonomous aspect.

In order to contribute to tackling the above mentioned issues, we propose a hybrid intrusion detection framework based on an active and incremental learning approach. The IDS includes mainly two modules, a model-based detector which is a deep neural network, and an unsupervised-based detector, namely, the KNN algorithm. The two modules cooperate in order to improve the performance of the IDS in an incremental fashion. Indeed, the input of the IDS is a flow of sliding-windows of the network traffic where the results of a specific window are leveraged for the analysis of the future windows. Each window is analyzed in order to detect known and unknown attacks with the previously-mentioned modules, but also in order to retrain and to improve the DNN-based detector for the next windows. To this purpose,  a query is made to the expert to label some data examples, thus the system is considered as an active learner with which the expert's effort in terms of labeling is minimized. The examples to label are selected with the presented query function which is based on two sampling strategies. The first strategy relies on the classification uncertainty which is the DNN detection uncertainty. The second sampling strategy consists in selecting the eventually unknown attacks detected by the KNN module. Once the sampled data are labeled, they are saved to a pool of labeled data obtained from the previous windows, and then used to retrain the DNN to better detect the known attacks but also and more importantly to teach it and update it on how to detect the newly discovered unseen attacks.

The main contributions of our work can be summarized as follows:

\begin{itemize}
\item A hybrid intrusion detection framework for known and unknown attack detection is proposed.
\item Active learning is exploited in order to take a step towards an autonomous IDS by minimizing the intervention of security experts in terms of data labeling.
\item A hybrid query function is proposed in order to improve the known attacks detection and to update the IDS with the newly-discovered attacks, making it, in this way, adaptive.
\item The IDS analyzes the network traffic through a sliding-window, it is thus incremental and more adapted to the changing network traffic.
\item Appropriate experiments are conducted in order to evaluate the proposed IDS in terms of  new attack detection, incremental learning but also in terms of data labeling cost minimization. 
\end{itemize}

The rest of the paper is organized as follows. The related works are reviewed in section \ref{rw}.  Section \ref{ids} presents the details of the proposed intrusion detection framework. In Section \ref{exp}, the evaluation of the framework is conducted. Finally, Section \ref{conclusion} concludes the article. 

\section{Related work}
\label{rw}

Intrusion detection systems are numerous and can be differentiated with regard to their functioning and their methods. Nowadays, the majority of IDSes used in the industry are signature-based, such as Suricata \cite{R2.1} and SNORT \cite{R2.2}. With these systems, security experts design a signature for each new attack after its occurrence and subsequently update the database of the IDS. Therefore, an activity is flagged as anomalous only if it presents a pattern that matches a known attack signature. As it can be concluded, these systems do not cope with previously unseen attacks.

Due to the drawbacks of the signature-based IDSes, the research community has explored more intelligent techniques, notably machine learning. With these techniques, the historical behavior, either normal or malicious behavior, is modeled and deployed. Subsequently, when a new activity is compared to the learned model, it is considered anomalous if it fits the malicious behavior model or if it deviates from the normal behavior model. Some examples of this category can be found in \cite{R2.3}-\cite{R2.7}. These IDSes are also debatable because they rely heavily on prior knowledge about what constitutes the normal or malicious behavior; i.e. labeled data. Additionally, they suffer from an important rate of false alarms, especially the normal behavior models. And the malicious behavior models fail to detect unknown attacks.
 
Two other categories of IDSes that are gaining more interest are the unsupervised and the hybrid. Unsupervised IDSes rely on completely unsupervised techniques such as outlier and clustering-based anomaly detection algorithms. These techniques work without a training phase and distinguish between normal and anomalous instances without the need for data labels. Moreover, the major advantage of these techniques is their ability to detect new types of anomalies. Unsupervised intrusion detection approaches are presented in \cite{R2.72}\cite{R2.71}-\cite{R2.10}. However, these detection strategies suffer from a low detection accuracy. Hybrid IDSes combine above-mentioned approaches in order to achieve a better performance in terms of detection rate and false alarms rate.  Some hybrid solutions can be found in \cite{R2.11}-\cite{R2.15}.

The above discussed methodologies could help in automatizing the intrusion detection, however, an optimal IDS should be scalable, adaptive and autonomous. Some efforts have been done in this direction.
 A promising approach which helps in taking a step towards a near-autonomous detection is active learning. The strategy has been initiated in \cite{R2.16}-\cite{R2.21}. In \cite{R2.16}, the authors have presented a general AL framework for intrusion detection.  The support vector machines (SVM) is used as the classifier. The points closest to the SVM hyperplane are considered as those of which the classifier is the most uncertain. However, a balanced query function has also been suggested in order to alleviate the skewness of the labeled data.  The authors in \cite{R2.17} have proposed a supervised solution based on the Transductive Confidence Machines for K-Nearest Neighbors (TCM-KNN) in an active learning approach. The P-values resulting from the TCM-KNN algorithm are used to define the uncertainty based query function.  A variant of the SVM algorithm, namely, the support vector domain description (SVDD), is used in \cite{R2.18} in an active approach. Herein, the query function relies on selecting both labeled and unlabeled data, it selects examples that are close to the boundary (margin strategy) but also the examples which lie in potentially anomalous and variable clusters. In \cite{R2.19}, the authors have proposed a framework that leverages both the co-training and the active learning concepts. The feature set is divided into two subsets to be exploited in the co-training strategy. A naive bayes classifier is used as the learner. An entropy-based query function and a nearest neighbor based method for rare category detection are used for the active learning approach. In the experiments, the authors have demonstrated how the proposed strategy could enhance the reduction of the false positive rate. Recently, the AL along with transfer learning have been used for intrusion detection in \cite{R2.20}.
 
The active learning approach helps in reducing the labeling effort for the security experts. However, it has its limits, more precisely, it won't be able to detect new attacks in case their pattern differs significantly from known attacks. Indeed the query function, aims mainly to improve the distinction between the attacks known to the model and the normal traffic. In the reviewed articles, almost none of them have attempted to "teach" the classifier how to detect new attacks in order to make it adaptive and scalable. The work in \cite{R2.19} is the only one suggesting the detection of rare events, however, the authors didn't emphasize on the unknown attacks detection. To the best of our knowledge, our work is the first to use active learning in an incremental fashion in order to build an adaptive IDS in terms of unknown attacks detection and minimization of the data labeling effort. 

\section{Proposed intrusion detection System}
\label{ids}
The concepts and the constituent parts of the IDS are presented and detailed in this section.  

\subsection{Data pre-processing}
Network traffic is aggregated into flows which are then described with a set of different statistics generated through the feature creation step. However, the created dataset might include missing values. The replacement of these entries may cause the data to be misleading, thus, we chose to delete them.

The features values might belong to different ranges, lowering in this way the performance of the  ML model. Standardization allows transforming the data features so as they have one common scale, it is defined as in the following formula:

\begin{equation}
x'=\frac{(x- \mu)}{\sigma}
\end{equation}
where $x$ is the value in the considered feature of a specific data point,  $\mu$ represents the mean and $\sigma$ the standard deviation of the numeric values of the considered feature.

\subsection{Detection modules}
\label{model}
Our IDS includes two detectors, a supervised module which is a DNN included in the incremental and active learning and an unsupervised detector which is the KNN algorithm.
\subsubsection{The supervised detector }

The architecture of the DNN, inspired from \cite{R2.19.1}, includes an input layer with a number of neurons adapted to the analyzed data, four  hidden layers, the first layer has 256 neurons, the second layer has 128 neurons, the third one has 64 neurons, while the last hidden layer includes 32 neurons. All of these layers are fully-connected and use the same activation function which is the Relu function. The last layer is the output of the network, since we are performing a binary classification, a 2 neurons layer with the Softmax activation function is used. For the model training, two elements are necessary, namely, a loss function and an optimization algorithm. The cross entropy loss function and the Adam optimizer \cite{R2.19.2} have been used for training the model and learning its weights. The intrusion detection data are imbalanced, thus the DNN is a weighted DNN, it consists in applying  a large error weighting to those examples in the minority class during the training step \cite{R2.19.3}.

\subsubsection{The unsupervised module}

We use the KNN algorithm to help in detecting new attacks. With it, an outlierness score is assigned to each data sample in the dataset where outliers tend to have higher scores. The score of outlierness is defined as the distance of the given data sample to its $K^{th}$ nearest neighbor.

\subsection{Query function}

\begin{algorithm}[h]
\caption{Uncertainty and KNN-based Query Function  }
\label{alg1}
\begin{algorithmic}
\State \textbf{Input:}
\State $X$: Network traffic data
\State $os$ : Outlierness score of each sample in $X$
\State $n$ : Number of samples to select for labeling
\State $\mathcal{M}$ : The ML model

\State \textbf{Output:}
\State \textbf{$\mathcal{L}$} : Labeled data
\State \textbf{Begin}
\State \hspace{2 mm} $	i1\gets$ top $n/2$ samples in $os$ \Comment{with the highest degree of outlierness}
\State \hspace{2 mm} $	u\gets$ Uncertainty($X$, $\mathcal{M}$) \Comment{based on Equation \ref{eq1}}
\State \hspace{2 mm} $	i2\gets$ top $n/2$ samples in $u$ \Comment{with the highest uncertainty}

\State \hspace{2 mm} $	I\gets$ $i1 \cup i2$

\State \hspace{2 mm} Query the oracle for labeling the data samples in $I$
\State \hspace{2 mm} Add the labeled data to $\mathcal{L}$

\State \textbf{End}
\end{algorithmic}
\end{algorithm}

As shown with the pseudocode in Algorithm \ref{alg1}, our query function relies on the uncertainty-based sampling, but also on KNN. The uncertainty-based sampling is defined in Equation \ref{eq1}, where x is the data point to be predicted and $\hat{x}$ is the most likely prediction. It is used to request the labels of the data points for which the DNN is the least confident about.

\begin{equation}
\label{eq1}
Uncertainty(x) = 1 - P(\hat{x}|x)
\end{equation}
 
On the other hand, the results of the KNN are exploited to select unknown intrusions to integrate in the active learning. Indeed, the data points for which the algorithm assigns a high score of outlierness are presented to the oracle for labeling and subsequently added to the labeled data pool $\mathcal{L}$ to retrain the DNN. This sampling will allow the neural network to learn how to detect new attacks. The number of samples to select is determined with a parameter $n$.

\begin{figure}[h]
\center \includegraphics[scale=0.45, width=85mm]{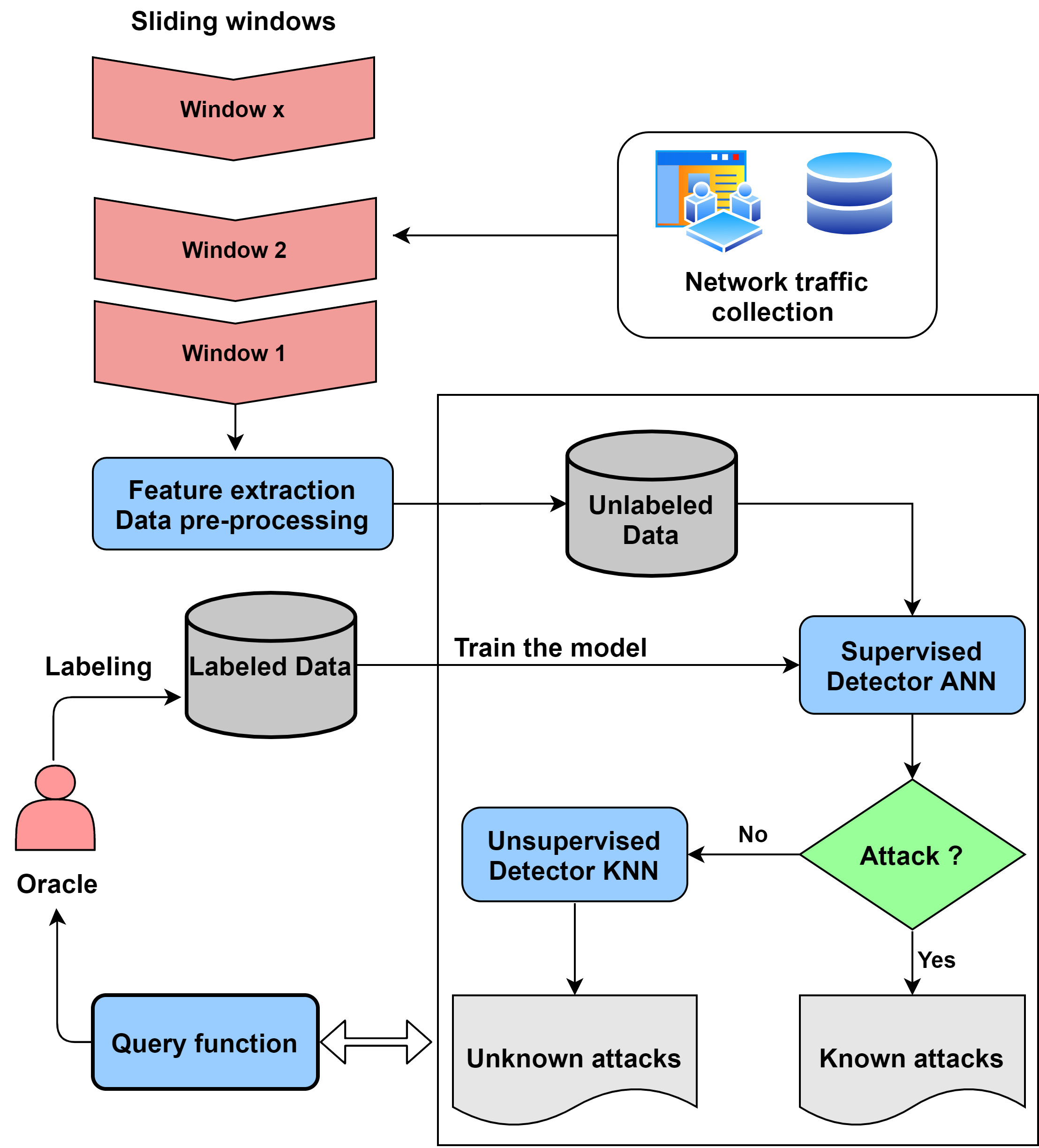}
\caption{The general functioning of the proposed IDS}
\label{fig2}
\end{figure}

\subsection{The general functioning of the IDS}

The flowchart in Figure \ref{fig2} shows the general functioning of the proposed intrusion detection framework. The system includes mainly two modules for detection, the supervised DNN which is integrated in the active learning as the model to "teach", and the unsupervised KNN detector which allows the detection of unknown attacks and their integration in the query strategy so as to allow the DNN to learn the detection of these new attacks and making it adaptive.

The proposed IDS is considered as incremental where the data are processed based on a sliding window (SW) of size s. Consider $SW_t$ the sliding window presented to the IDS at time t. After the pre-processing step, the DNN-based detector is used on the window to find the attacks known to the algorithm. Subsequently, the DNN results are processed in order to detect other intrusions, especially unknown ones, through the KNN algorithm, but also in order to select the data instances of which the DNN is the least confident by using the uncertainty sampling function. Results of the  query function are checked by the security expert through the labeling process, the labeled data are then added to a dedicated pool and used to retrain the DNN model. The new model is later used to analyze window $SW_{t+1}$ and the whole detection process is repeated again.

\begin{table*}[h]
\center
\caption{Summary of the learning data, of the  different SWs , and of the test data. The double checkmark (\checkmark \kern-0.6em\checkmark ) indicates that the attack is unknown so far}
%\begin{center}
\begin{tabular}{|p{1.5cm}|p{1cm}|p{1cm}|p{1.2cm}|p{1.2cm}|p{1.2cm}|p{1.2cm}|p{1cm}|p{1cm}|p{0.9cm}|p{1.1cm}|}
%\begin{tabular}{|p|p|p|p|p|p|p|}
\hline
\textbf{ }&\multicolumn{10}{|c|}{\textbf{Network traffic type }} \\
\cline{2-11} 
\textbf{Data} & \textbf{\textit{Normal}}& \textbf{\textit{Hulk}}& \textbf{\textit{GoldenEye}}& \textbf{\textit{Slowhttptest}} & \textbf{\textit{Slowloris}}& \textbf{\textit{Heartbleed}}&
\textbf{\textit{Brute Force}}&
\textbf{\textit{SQL Injection}}&
\textbf{\textit{XSS}}&
\textbf{\textit{Port Scan}} \\

\hline
SW1(labeled) &  \checkmark & - & \checkmark & -& - & \checkmark& - & \checkmark & - & - \\

\hline
SW2-SW15& \checkmark & - & \checkmark  & - & - &  & - & - & - & - \\
\hline
SW16-SW26& .\checkmark & - & - & - & - & - & - &- & -& \checkmark \kern-0.6em\checkmark  \\
\hline
SW27-SW30& \checkmark & - & \checkmark  & - & \checkmark \kern-0.6em\checkmark  & - &\checkmark \kern-0.6em\checkmark  &- &- & - \\
\hline
SW31-SW36& \checkmark & - & \checkmark & \checkmark \kern-0.6em\checkmark  & \checkmark  & - & - & - & -& - \\
\hline

SW37-SW40&\checkmark  & - & \checkmark &  \checkmark  & \checkmark & - & - & - & \checkmark \kern-0.6em\checkmark  & - \\
\hline
SW41-SW52& \checkmark & \checkmark \kern-0.6em\checkmark  & \checkmark  & \checkmark & \checkmark & - & - & - & - & -\\
\hline

SW53-SW56& \checkmark & \checkmark & -  & \checkmark & \checkmark & - & - & - & - & -\\
\hline
SW57-SW60& \checkmark & - & - & \checkmark & -& - & - & - & - &  \checkmark \\
\hline
Total (SWs)& 254202& 19723&  3498& 2358&2579& 8& 238 & 15 & 233 &  26145\\
\hline
Test data& 72500 & 25000 & 3050 & 1100   & 1200& 3& 450 & 6 & 190 & 17500 \\
%copy& More table copy$^{\mathrm{a}}$& &  \\
\hline

\end{tabular}
\label{tab1}
%\end{center}
\end{table*}

\section{Experimental evaluation}

\label{exp}
In order to evaluate the performance of the proposed IDS, a set of experiments has been conducted as presented in this section.

\subsection{Dataset}

The CICIDS2017 dataset \cite{R4.3} which is available online at \cite{R4.4} has been explored in our experiments. The dataset has been captured from July 3rd to July 7th, 2017. The data are described with 77 traffic features and contain a realistic background traffic and the most up-to-date common attacks. The benign background traffic includes the abstract behaviour of 25 users based on the HTTP, HTTPS, FTP, SSH, and email protocols. In our case, in addition to benign traffic, a set of different attacks was selected, namely, DoS attacks (Hulk, GoldenEye, Slowloris and Slowhttptest), Port scan, Heartbleed attacks and Web attacks (Brute Force, SQL Injection and XSS). 

The objective from our experiments is to test (i) the ability of the IDS  to achieve a good performance with a minimum amount of labeled data selected through the active learning and (ii) its ability to improve and adapt its performance through the incremental learning of unknown attacks. To this purpose, the dataset was exploited in a way that allows simulating a real life scenario. We have divided it into two sets, a training or learning set of 300000 samples and a test set of 121000 samples, an equivalent of a split of around 70 - 30 \% for the learning-test sets.
 
 Since, in our case, the proposed solution is incremental, the training set was split into 60 SWs of size $s$ equal to 5000 data instances. The first window (SW1) is completely labeled to allow a first training of the neural network. In the remaining windows, in order to simulate unseen attack occurrence, some types of attacks are included gradually. The learning data, i.e. the sliding windows, and the test data are summarized in Table \ref{tab1}.
 
\subsection{Experimental setup}

The implementation of the proposed scheme was achieved by using two Python packages, namely, modAL \cite{R4.1} and Pyod \cite{R4.2}. The neural network in our case is built by using Keras with the characteristics mentioned in Subsection \ref{model}. Our query function has been implemented and used along with the DNN.

The training of the DNN model by using the labeled data is performed with batch size of 32 and with the early stopping option. Since our DNN is a weighted neural network, the weights 1 and 10 are used for the normal class and attack class, respectively. The Adam optimizer is applied with the $learning-rate$ equal to 0.001 and the exponential decay rates $\beta 1$ = 0.9 and $\beta 2$ = 0.999. The K-nearest neighbor algorithm was used with a number of neighbors $K$ equal to 100.

The effect of the parameter $n$ which represents the number of instances to query for labeling is examined with experiments where $n$ takes different values.

\subsection{Evaluation measure}
In order to evaluate the performance of the IDS, the Area Under the ROC Curve (AUC) is used. The Receiver operating characteristic (ROC) curve is obtainted with a plot of the true positive rate (TPR) as a function the false positive rate (FPR) at various threshold settings. TPR and FPR are defined as in the following.

\begin{equation}
TPR=TP/(FN+TP)
\end{equation}
\begin{equation}
FPR=FP/(FP+TN)
\end{equation}
where FP are the false positives, TP are the true positives, TN are the true negatives and FN are the false negatives.

\subsection{Results and discussion}
The results obtained on the test set and on the different sliding windows  are presented and discussed in the following.

\begin{figure}[htbp] 
\includegraphics[scale=0.65]{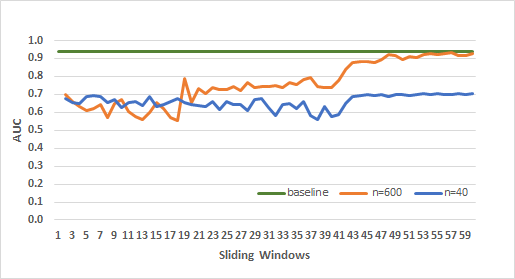}
\caption{Results of the incremental and active learning-based IDS on the test set with different number of queried samples (n) in each SW, the baseline is obtained by using full training set}
\label{fig2.0}
\end{figure}

\begin{figure}[htbp] 
\includegraphics[scale=0.65]{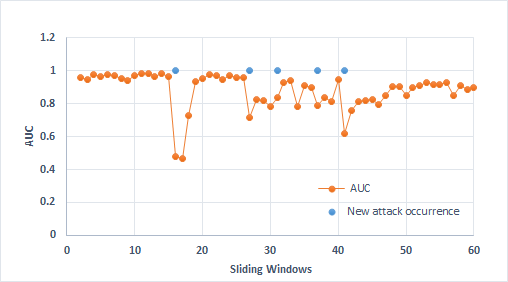}
\caption{Results of the incremental and active learning-based IDS on the different sliding windows} \label{fig2.1}
\end{figure}

Figure \ref{fig2.0} presents the AUC of the incremental active learning approach on the test set. The results are obtained by training the deep neural network with a different number of queried samples $n$ for labeling in each sliding window, which are compared to the baseline obtained by using the full training data. The parameter $n$ was set to  40 and 600 , an equivalent to 2\% and 13\% of the full training set, respectively, and including the labeled window SW1. From the plots, we can notice  the dependence of the ML model on the number of labeled data, i.e. the more labeled data we provide, the better the DNN performance. However, we can also see that labeling only 600 samples from each window, i.e. 13\% of the full training data allows us to achieve a performance as good as when using the full training set, thus lowering considerably the labeling cost and effort.

Since the test set includes all the attack types, the results from Figure \ref{fig2.0} allow us to demonstrate the incremental aspect of the detection approach. As can be seen, at the very beginning, the detection performance is low, because at this stage, the sliding windows presented to the system contained only three types of attacks, namely, the GoldenEye,  the Heartbleed and the SQL Injection attacks, the system is thus unable to detect the unseen attacks in the test set. After the introduction of new attacks, a clear increase of the performance can be noticed, for instance, the AUC goes from 0.77 to around 0.9 from SW40 to SW48, indeed, herein the Hulk attack was introduced, and the system started learning to detect it.

However, it can be noticed that at the very beginning, i.e. from SW2-SW17, when using only $n$=40, the obtained AUC is superior to the one obtained with $n=600$. This can be explained by the good performance of the DNN on these windows as can be seen in Figure \ref{fig2.1}. Indeed, at this stage, the GoldenEye attack is known to the detector, thus it is well detected and few attacks remain undetected in the sliding windows and the queried samples for labeling are therefore mostly benign, therefore the less queried benign samples, the better the detection, especially that the test set is imbalanced.

The adaptive aspect of the detection system can be seen in Figure\ref{fig2.1}. The plot represents the AUC obtained by using $n=600$ on the different sliding windows where unseen attacks appear gradually. The blue dots indicate the sliding windows where unseen attacks occur.

From the plot, we can see a good performance of the DNN detector, however, we notice a sharp drop in the AUC measure each time a new attack is introduced to the system, indicating the failure of the neural network to detect the unseen attacks. For instance, when the Port Scan intrusion occurs in sliding window 16, the AUC drops from 0.97 to less than 0.50. However, the system subsequently resumes its performance as the new attacks are detected by the query function and tough to the DNN through the active learning process. Indeed the Port Scan attacks present from SW19 to SW26 have been well detected with an AUC around 0.96. As such, we can confirm that results of the current traffic window are leveraged to improve the performance of the detector for the future windows by helping it to adapt to the new traffic and attacks.

\section{Conclusion and future work}
\label{conclusion}
In this article, an intrusion detection system based on incremental active learning is presented. The proposed approach is considered as adaptive given that it learns how to detect both known and unknown attacks through the uncertainty and KNN-based query function for interactive labeling. It is also incremental due to the fact that the detection model performance improves with time. Furthermore, a step towards an autonomous detection is taken since the active learning helps in reducing the efforts of the experts as it allows a significant reduction of the amount of data to label. Indeed, the conducted experiments show that the system performance by using only 13\% of the data is as good as when the full data are used.

As future work, we intend to improve the framework and compare it with state-of-the-art solutions. We intend to explore more complex deep learning architectures and thus using more data and conducting more experiments to understand the effect of the different parameters, especially the size of the sliding window. In addition to the detection enhancement, the deep learning model will be able to perform a multi-class classification where the model is able to distinguish the different types of attacks helping in this way the expert to interpret the detection results. We also plan to improve and evaluate the unsupervised detector by potentially exploring outlier ensembles where results of a set of different algorithms will be considered for a more accurate detection of the unknown attacks. 
Another aspect which might be interesting to explore is the attacks explanation \cite{R4.5}, herein, the aim is to help the user to understand the malicious traffic by providing more information on what characterizes the detected attacks, especially the new ones.

\section*{Acknowledgment}

This work was funded by the National Natural Science Foundation of China, Grant number 61272420 and 61472189.

\end{document}